\documentclass[reprint, amsmath,amssymb, aps,superscriptaddress,longbibliography]{revtex4-1}
\usepackage[dvips,letterpaper,margin=0.8in,bottom=1in]{geometry}
\usepackage{amsmath}
\usepackage{amssymb}
\usepackage{graphicx}
\usepackage[normalem]{ulem}
\usepackage{float}
\usepackage{comment}
\usepackage{xcolor}
\usepackage{tikz}
\usepackage{lipsum}
\usepackage{hyperref}
\usepackage[utf8]{inputenc}

\newcommand{\citeasnoun}[1]{Ref.~\cite{#1}}
\newcommand{\figref}[1]{Fig.~\ref{fig:#1}}
\newcommand{\figsref}[2]{Figs.~\ref{fig:#1}~and~\ref{fig:#2}}
\newcommand{\figrefbegin}[1]{Figure~\ref{fig:#1}}

\renewcommand{\eqref}[1]{Eq.~(\ref{eq:#1})}
\newcommand{\eqrefbegin}[1]{Equation~(\ref{eq:#1})}

\newcommand{\eqssrefbegin}[2]{Equations~(\ref{eq:#1}--\ref{eq:#2})}

\newcommand{\eqssref}[2]{Eqs.~(\ref{eq:#1}--\ref{eq:#2})}

\newcommand{\KET}[1]{|\!\left.{#1}\right>}

\newcommand{\Scale}[2][4]{\scalebox{#1}{$#2$}}

\makeatletter
\newcommand*{\encircled}[1]{\relax\ifmmode\mathpalette\@encircled@math{#1}\else\@encircled{#1}\fi}
\newcommand*{\@encircled@math}[2]{\@encircled{$\m@th#1#2$}}
\newcommand*{\@encircled}[1]{%
  \tikz[baseline,anchor=base]{\node[draw,circle,outer sep=0pt,inner sep=.2ex] {#1};}}
\makeatother

\begin{document}

\title{Boosting photonic  quantum computation with moderate nonlinearity}
%{Single-photon phase nonlinearity can help  quantum computation}

\author{A. Pick}\thanks{adi.pick@weizmann.ac.il}
\affiliation{Department of Chemical Physics, Weizmann Institute of Science, Rehovot
76100, Israel}
\author{E. S. Matekole}
\affiliation{Department of Physics and Astronomy,
Louisiana State University, Baton Rouge, LA 70803, USA}
\author{Z. Aqua} 
\affiliation{Department of Chemical Physics, Weizmann Institute of Science, Rehovot
76100, Israel}
\author{G. Guendelman }
\affiliation{Department of Chemical Physics, Weizmann Institute of Science, Rehovot
76100, Israel}
\author{O. Firstenberg}
\affiliation{Department of Physics, Weizmann Institute of Science, Rehovot
76100, Israel}
\author{J. P. Dowling}\thanks{Deceased 5 June, 2020.}
\affiliation{Department of Physics and Astronomy,
Louisiana State University, Baton Rouge, LA 70803, USA}
\author{B. Dayan}
\affiliation{Department of Chemical Physics, Weizmann Institute of Science, Rehovot
76100, Israel}

%----------------------------------------------------------------------
\begin{abstract}
Photonic measurement-based quantum computation (MBQC) is a promising route towards fault-tolerant universal quantum computing. A central challenge in this effort  is the huge overhead in the resources required for the construction of large photonic clusters using probabilistic linear-optics gates. Although strong single-photon nonlinearity ideally enables deterministic construction of such clusters, it is challenging to realise in a scalable way. Here we explore the prospects of using moderate nonlinearity (with conditional  phase shifts smaller than $\pi$) to boost photonic quantum computing and significantly reduce its resources overhead. The key element in our scheme is a nonlinear router that preferentially directs photonic wavepackets to different output ports depending on their intensity. As a relevant example, we  analyze the nonlinearity provided by Rydberg blockade in atomic ensembles, in which the trade-off between the nonlinearity and the accompanying loss is well understood. We present protocols for efficient Bell measurement and GHZ-state preparation -- both key elements in the construction of cluster states, as well as for the CNOT gate and quantum factorization. Given the large number of entangling operations involved in fault-tolerant MBQC, the increase in success probability provided by our protocols already at moderate nonlinearities can result in a dramatic reduction in the required  resources. 
\end{abstract}
%----------------------------------------------------------------------

\maketitle
%----------------------------------------------------------------------
Photonic quantum computation is a leading platform in the effort towards fault-tolerant universal quantum computers~\cite{kok2007linear,o2009photonic,rudolph2017optimistic,Bartolucci2021fusion}. It combines the paradigm of measurement-based  quantum computation (MBQC)~\cite{gottesman1999demonstrating,raussendorf2001one,knill2001scheme,nielsen2003quantum,verstraete2004valence,leung2004quantum,briegel2009measurement}, where the computation is carried out by applying  a sequence of  measurements to entangled cluster states~\cite{briegel2001persistent,hein2004multiparty,nielsen2004optical} with topological quantum error correction~\cite{gottesman1997stabilizer,bravyi2005universal,varnava2006loss,dawson2006noise,raussendorf2007topological,fowler2012surface}.  In particular, the promise of all-optical photonic quantum computation  with discrete variables~\footnote{Our work focuses on discrete-variable quantum computation. While large clusters of squeezed states have been demonstrated in continuous-variable quantum computation~\cite{yokoyama2013ultra,chen2014experimental,reimer2016generation,asavanant2019generation}, the latter approach faces other challenges that currently hinder its scalability, including the generation of Gottesman-Kitaev-Preskill (GKP) states~\cite{gottesman2001encoding}.} lies in the  ability to entangle single  photons into graphs and clusters using only  linear-optics probabilistic operations~\cite{zeilinger1997three,browne2005resource,lu2007experimental,kieling2007percolation,wilde2007alternate,shadbolt2012generating,wang2016experimental,istrati2020sequential}. The price, however, is a huge overhead: constructing a cluster of  $10^7$ photons  (corresponding to $\sim1000$ logical qubits, assuming $10^4\times$ redundancy for error correction) with probabilistic  gates may require $10^{12}$ input single photons~\cite{li2015resource}.

One approach to tackle  this challenge is efficient and  strong interaction with single quantum emitters, such as atoms, ions, or quantum dots. 
Such coupling ideally enables deterministic construction of cluster states either by generation of a stream of entangled photons ~\cite{schon2007sequential,lindner2009proposal,schwartz2016deterministic,pichler2017universal}, or by entangling single photons via photon-atom quantum gates~\cite{duan2004scalable,reiserer2014quantum,hacker2016photon,rosenblum2017analysis,bechler2018passive,borregaard2019quantum}. However, achieving strong interaction with single quantum emitters requires challenging optical structures, which are not straightforwardly scalable. A number of theoretical works explored the possibility of enhancing weak Kerr-type nonlinearities by classical driving fields to make them strong enough to support photonic quantum computation~\cite{paris2000optical,nemoto2004nearly,nemoto2005universal,munro2005weak,munro2005high,barrett2005symmetry,shapiro2006single,louis2007efficiencies,shapiro2007continuous,gea2010impossibility}. 
To be precise,  we define nonlinearity as ``strong'' if it can provide a conditional phase shift $\varphi=\pi$; namely, a difference of $\pi$ between twice the phase acquired by a single photon in a mode, and the phase acquired by two photons in the same mode. 

In contrast to previous studies, here we  explore whether there is an intermediate regime between the linear optics and  strong nonlinearity regimes in which moderate nonlinearity can provide a practical advantage. The motivation is that since  fault-tolerant photonic quantum computation involves a large number of probabilistic gates~\cite{li2015resource}, even a small improvement in the success probability per gate could amount to a dramatic reduction in the required resources.
The reason for focusing on moderate nonlinearity is that attaining small phase shifts (e.g., $\pi/10$) typically requires significantly less resources than attaining $\pi$, and, accordingly, is accompanied by much lower costs, such as, in particular, photon loss, which is the dominant fault mechanism in photonic qubits. 
Although advanced protocols for photonic quantum computing can tolerate up to 50$\%$ overall photon loss (being an inherently detectable error in most cases)~\cite{varnava2006loss,stace2009thresholds,barrett2010fault,Bartolucci2021fusion}, the large number of elements and operations involved makes reaching this level challenging nonetheless. Given the progress in photonic technologies, including large scale on-chip single-photon detection capabilities with superconducting nanowires~\cite{marsili2013detecting,reddy2020superconducting}, the additional loss induced by the nonlinear medium itself will likely become the dominant one. We therefore wish to quantify the trade-off offered by introducing nonlinear elements to the task of graph-states construction, namely increasing the success probability of each operation at the price of significant additional loss.  

In order to take  loss into account as accurately as possible, we consider below the very relevant and well-studied platform of electromagnetically-induced transparency (EIT~\cite{boller1991observation}) with Rydberg atoms (Rydberg-EIT)~\cite{distante2017storing,tiarks2019photon}. In these systems, the loss grows quadratically with the conditional phase shift $\varphi$  (for small $\varphi$), which is eventually limited by the physical parameters of the Rydberg ensemble~\cite{lahad2017induced}.

% ---------------------------------------------------------------------------
% FIGURE 1 
% ---------------------------------------------------------------------------
\begin{figure}[t]
\centering
                 \includegraphics[width=0.5\textwidth]{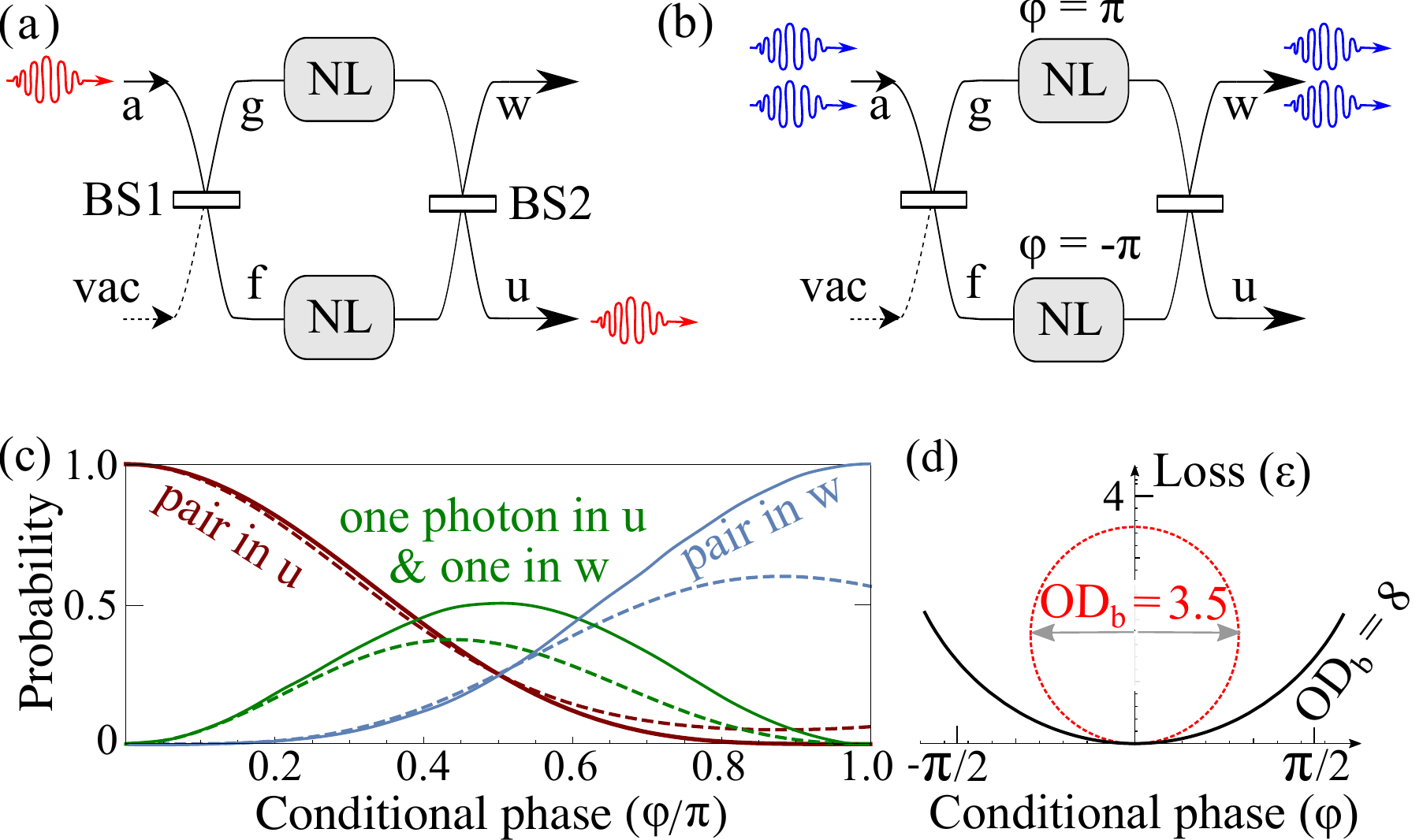}
                  \caption{\textbf{Nonlinear router:} (a) When a photon enters the interferometer in mode $a$, it leaves in  $u$. (b) The interferometer  contains  a nonlinear     medium (NL, gray) in both arms. Photon pairs traveling through  different arms acquire opposite conditional phases $\pm\varphi$ [\eqref{cross-phase-shift}]. When $\varphi = \pi$, a pair that enters in mode $a$ leaves in mode $w$. (c) Probabilities for  detection  outcomes for an incoming  pair in mode $a$ when $\varphi\in[0,\pi]$: both  photons  in $u$ (brown),  both in $w$ (blue),  and one   in each mode  (green). Solid lines show a loss-free model [\eqref{NL-MZI}].  Dashed lines include  nonlinear loss  in Rydberg-EIT systems [\eqref{MZI-with-loss}]. We assume that  the optical depth of the blockaded volume is   $\text{OD}_\text{b} = 30$, which  sets the loss probability $\tau(\varphi)$ via  \eqref{the-circle}, as shown in  (d).  The relation between the attenuation coefficient, $\varepsilon \equiv -\ln(1-\tau)$, and $\varphi$ is a circle~\cite{lahad2017induced},  plotted for $\text{OD}_\text{b} = 3.5$ and  8 (red and black).
}
                  \label{fig:nl-router}
  \end{figure}
% ---------------------------------------------------------------------------

\begin{figure}[t]
\centering
                 \includegraphics[width=0.5\textwidth]{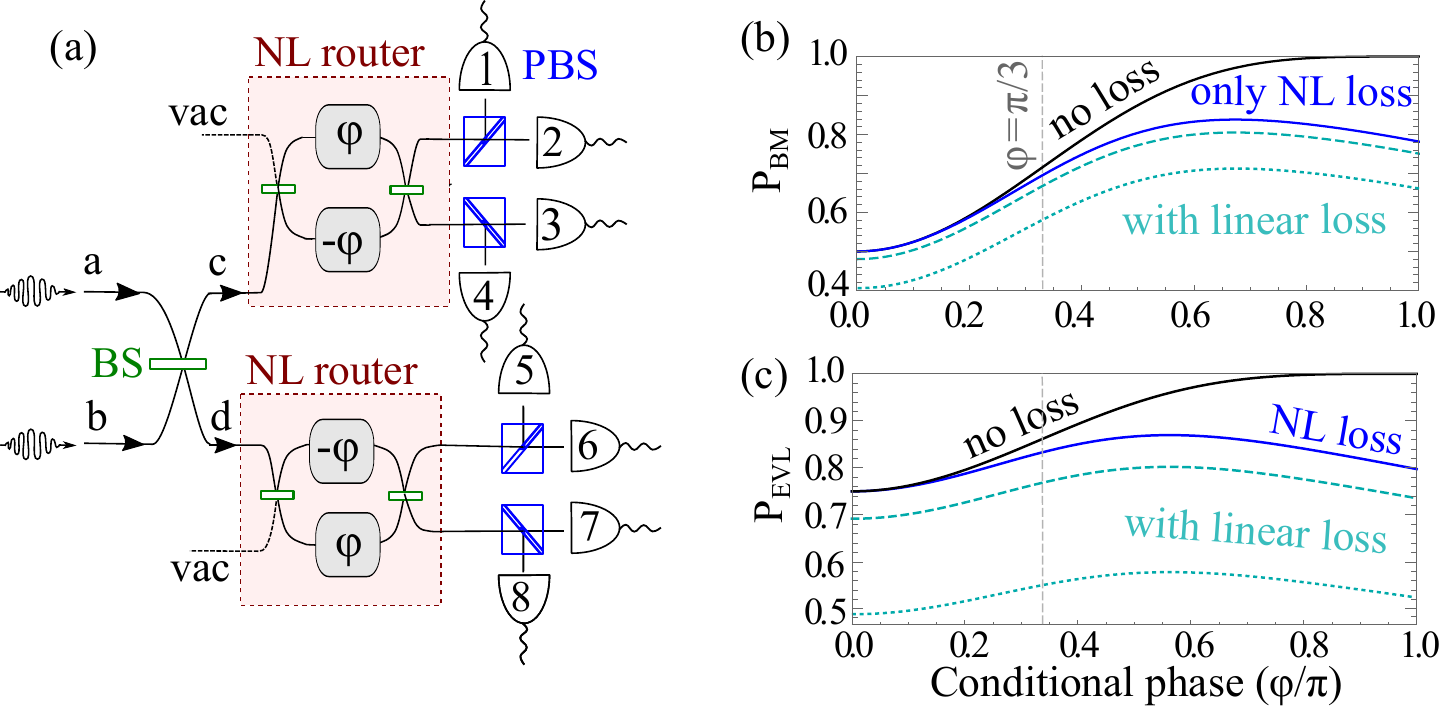}
                  \caption{\textbf{Nonlinear Bell measurement (BM):} 
Two photons are sent through a beam splitter (BS) and nonlinear routers before reaching polarizing beam splitters (PBS) followed by  detectors (1--8), which  measure their joint polarization state in the Bell basis.  While $\KET{\psi_\pm}$  produce distinct  final states,  $\KET{\phi_\pm}$ can only be distinguished with nonlinearity.
(b) Success probability of a BM, $P_\text{BM}$, as a function of $\varphi$. Black solid  curve neglects loss [\eqref{BM-success} with $\tau=0$ and $P_\text{DE} = 1$]. Blue solid curve includes only nonlinear loss,  assuming $\text{OD}_\text{b} = 30$, while the cyan dashed (dotted) curve includes also linear loss with $P_\text{DE} = 98\%$ ($90\%$). With $P_\text{DE} = 98\%$ and $\varphi = \pi/3$, $P_\text{BM}$ improves from $0.48$ to $0.67$. 
 (c) Success probability of the nonlinear enhancement of the Ewert-van Loock  protocol~\cite{ewert20143}, $P_\text{EVL}$ [\eqref{EVL-success}]  that includes two  ancillary qubits. Curves are labeled as in (b).
With $P_\text{DE} = 98\%$ and  $\varphi = \pi/3$,   $P_\text{EVL}$ improves from $0.69$ to $0.77$.
}
\label{fig:NL-BM}
  \end{figure}

The key element in our protocols is a nonlinear router~\cite{rosenblum2016extraction,tiecke2014nanophotonic,ralph2015photon}, realized by a Mach--Zehnder interferometer (MZI) with a nonlinear medium in both arms (\figref{nl-router}). The conditional phase-shift acquired by two-photon pulses (photon pairs) in the nonlinear media causes the router to preferentially direct single photon pulses to one port and photon pairs to another. 
We use the router to obtain nonlinearity-enhanced protocols for Bell-state  measurement (BM) and for  Greenberger--Horne--Zeilinger (GHZ) state preparation (\figsref{NL-BM}{NL-GHZ}), which are key building blocks in MBQC~\cite{gottesman1999demonstrating}. As an application, we use these elements to construct protocols for a CNOT  gate and  quantum factorization (\figref{NL-GC}).  By adding nonlinearity, our protocols outperform the linear methods, potentially reducing resource requirements for  fault-tolerant photonic quantum computing by  up to two orders of magnitude already at moderate nonlinearities.\\
 
% ---------------------------------------------------------------------------
\textit{Nonlinear router:}
% ---------------------------------------------------------------------------
As  shown  in~\figref{nl-router}(a), when single photons enter the MZI in mode $a$, they exit through port $u$, following the transformation rule~\cite{gerry2005introductory}:
\begin{gather}
a^\dagger \xrightarrow[]{\text{BS1}} 1/\sqrt{2}(f^\dagger + ig^\dagger) \xrightarrow[]{\text{BS2}} u^\dagger.
\label{eq:linear-MZI}
\end{gather}
The  MZI  contains a nonlinear atomic medium  that  induces opposite conditional phase shifts $\pm\varphi$ for photon pairs in each of its arms:
\begin{gather}
(f^\dagger)^2\rightarrow e^{i\varphi} (f^\dagger)^2
\quad,\quad
(g^\dagger)^2  \rightarrow e^{-i\varphi} (g^\dagger)^2.
\label{eq:cross-phase-shift}
\end{gather}
As a result, photon pairs  undergo the transformation
\begin{gather}
\Scale[0.9]{
\left(a^\dagger\right)^2 \!\! \xrightarrow[]{\text{BS1}}\! 
\tfrac{1}{2}\left(e^{i\tfrac{\varphi}{2}}f^\dagger + ie^{-i\tfrac{\varphi}{2}}g^\dagger\right)^2
\! \xrightarrow[]{\text{BS2}} \!
 \left(w^\dagger \sin\tfrac{\varphi}{2} + u^\dagger 
  \cos\tfrac{\varphi}{2}\right)^2}.
\label{eq:NL-MZI}
\end{gather}
Consequently, the probability of routing pairs to the second output mode $w$ increases monotonously with $\varphi$ [\figref{nl-router}(c)].  For $\varphi=\pi$,  the nonlinear router deterministically separates pairs to a different port than single photons [\figref{nl-router}(b)].

% ---------------------------------------------------------------------------
\textit{The effect of photon loss:}
% ---------------------------------------------------------------------------
Our schemes can be implemented with any nonlinear medium that  provides moderate conditional phase shifts, including tightly confined atomic ensembles (e.g., atom-cladded optical fibers~\cite{stern2013nanoscale,keil2016fifteen,kitching2018chip}, atom-filled hollow-core fibers\cite{bajcsy2009efficient,venkataraman2011few,venkataraman2013phase}, or optical traps~\cite{meng2018near,corzo2019waveguide,johnson2019observation}) and nonlinear fibers~\cite{matsuda2009observation}.
Here, we focus on  Rydberg-EIT systems to analyze effect of loss on our protocols. 
 We choose this platform  since it enables achieving moderate (and even large) conditional phase shifts~\cite{distante2017storing,tiarks2019photon}, since linear losses under EIT conditions can be made negligible~\cite{boller1991observation}, and since the phase-loss relation in this system is well understood \cite{lahad2017induced}.
In such systems, once a photon generates a Rydberg excitation, the energy levels of the surrounding atoms  (within the Rydberg-blockade radius) are shifted, violating the EIT conditions~\cite{gorshkov2011photon}. Consequently, any subsequent photon in this volume acquires a phase and suffers loss. 

We  model loss  as the annihilation of a photon (in    $f$ or $g$) and the creation of a photon in an undetected mode ($\ell$ or $k$). 
Denoting the absorption probability by $\tau$, a  photon pair in the atomic  medium follows the rule:
\begin{subequations}
 \begin{gather}
{f^\dagger}^2 \rightarrow 
\sqrt{1 - \tau}\,e^{i\varphi}{f^\dagger}^2   +
\sqrt{\tau}f^\dagger \ell^\dagger 
 \label{eq:possible-loss-f}
\\
{g^\dagger}^2 \rightarrow 
\sqrt{1 - \tau}\,e^{-i\varphi}{g^\dagger}^2   +
\sqrt{\tau}g^\dagger k^\dagger.
 \label{eq:possible-loss-g}
\end{gather} 
 \label{eq:possible-loss}
\end{subequations}
\hspace*{-3.6pt} The conditional phase shift $\varphi$ and  absorption coefficient $\varepsilon \equiv -\ln{(1-\tau)}$ follow the phase-loss circle~\cite{lahad2017induced}:
 \begin{gather}
\varphi^2 + (\tfrac{\varepsilon}{2} - \tfrac{\mathrm{OD}_\text{b}}{4})^2 =  (\tfrac{\mathrm{OD}_\text{b}}{4})^2,
\label{eq:the-circle}
\end{gather} 
where $\text{OD}_\text{b}$ is the optical depth of the blockade volume -- the core resource of nonlinearity in Rydberg-EIT systems~\footnote{There is a factor of 2 difference between our definition of  $\varepsilon$ and \citeasnoun{lahad2017induced}, since we refer to loss of modal amplitude and \citeasnoun{lahad2017induced} refers to intensity loss.}. As evident from \eqref{the-circle} and \figref{nl-router}(d), high $\text{OD}_\text{b}$ enables large conditional phase shifts with low loss. In particular, $\mathrm{OD}_\text{b} > 4\pi$ is required  for  $\varphi=\pi$. The value $\text{OD}_\text{b} = 13$ has already been reached experimentally~\cite{tresp2016single}, and higher values have been predicted~\cite{baur2014single,gaj2014molecular}. Additionally, high total $\text{OD}$ can be utilized to generate an effective cavity in the atomic medium with finesses $F\sim(\text{OD}/2)^{0.4}$~\cite{lahad2017induced}, leading to a phase-loss circle whose radius is  $F$ times larger, thereby enabling an effective $\text{OD}_\text{b}$ of $100$ or more.  Using   Eqs.~(1, 4, 5), we obtain
\begin{gather}
\Scale[0.89]{{a^\dagger}^2   \rightarrow
\tfrac{\sqrt{1-\tau(\varphi)}}{2}\left[\cos\varphi
\left({w^\dagger}^2 \! -{u^\dagger}^2  \right) - 2\sin\varphi (w^\dagger u^\dagger) \right]}
\nonumber\\
\Scale[0.89]{
\!-\tfrac{1}{2}
\left({w^\dagger}^2  +{u^\dagger}^2\right)+\tfrac{\sqrt{\tau(\varphi)}}{2\sqrt{2}}
  \left[(w^\dagger+iu^\dagger)\ell^\dagger - (u^\dagger+iw^\dagger)k^\dagger\right]\!.}
  \label{eq:MZI-with-loss}
\end{gather}
From  \eqref{MZI-with-loss}, we calculate the probabilities for different outcomes of the MZI when including loss [\figref{nl-router}(c)].

\eqssrefbegin{the-circle}{MZI-with-loss} assume that the photon wavelength is tuned
exactly on the EIT resonance,  where  the single-photon phase shift and loss are ideally zero. 
In the supplementary material (SM), we analyze the possibility of detuning from the EIT resonance to include the effect of single-photon phase shift and loss in the atomic medium. We find that the performance of our protocols is relatively unaffected by these processes.

% ---------------------------------------------------------------------------
% FIGURE 2
% ---------------------------------------------------------------------------
% ---------------------------------------------------------------------------

% ---------------------------------------------------------------------------
\textit{Nonlinear Bell-state measurement (BM):}
% ---------------------------------------------------------------------------
As a first application, we use the nonlinear router to improve linear BMs. 
 We begin this section by revisiting the traditional linear BM protocol. Then, we present our nonlinear protocol and discuss its key features in the limit of strong nonlinearity and negligible loss. Finally, we discuss moderate nonlinearity and non-negligible loss.

A  photonic qubit is defined as a single excitation in an arbitrary coherent combination of two non-overlapping optical modes. For convenience, here we use the linear polarization basis, yet our analysis is generally applicable to  other choices of optical modes, including dual-rail and   time-bin qubits. In this basis, the photonic two-qubit Bell states are $\KET{\psi_\pm} = \tfrac{1}{\sqrt{2}}(a_H^\dagger b_V^\dagger \pm a_V^\dagger b_H^\dagger)\KET{\text{vac}}$ and  $\KET{\phi_\pm} = \tfrac{1}{\sqrt{2}}(a_H^\dagger b_H^\dagger \pm a_V^\dagger b_V^\dagger)\KET{\text{vac}}$, where $x_{k}^\dagger$ denotes the creation operator of  a photon in mode $x$ with polarization $k$ operating on  the vacuum state $\KET{\text{vac}}$.
In a linear-optics BM,  photons are sent through a  balanced beam splitter  (BS)~\cite{braunstein1995measurement}.
 Applying the BS  transformation  $a_k^\dagger\rightarrow\tfrac{1}{\sqrt{2}}(d_k^\dagger +ic_k^\dagger)$ and $b_k^\dagger\rightarrow \tfrac{1}{\sqrt{2}}(c_k^\dagger +id_k^\dagger)$, one finds 
\begin{subequations}
\begin{gather}
\KET{\psi_-} \xrightarrow[]{\text{BS}} \tfrac{1}{\sqrt{2}}(d_H^\dagger c_V^\dagger - c_H^\dagger d_V^\dagger)\KET{\text{vac}},\\
\KET{\psi_+} \xrightarrow[]{\text{BS}} \tfrac{i}{\sqrt{2}}(d_H^\dagger d_V^\dagger + c_H^\dagger c_V^\dagger)\KET{\text{vac}},\\
\Scale[0.9]{\KET{\phi_\pm} \xrightarrow[]{\text{BS}}  \tfrac{i}{2\sqrt{2}}[(d_H^\dagger)^2+(c_H^\dagger)^2 \pm ((d_V^\dagger)^2+(c_V^\dagger)^2 ) ]\KET{\text{vac}}.}
\label{eq:phi-states}
\end{gather}
\label{eq:BS1-transformation}
\end{subequations}
\hspace*{-3.8pt}The states $\KET{\psi_\pm}$  lead to distinguishable outcomes. While $\KET{\psi_-}$ produces one photon in $c_k$ and one in $d_k$, the state  $\KET{\psi_+}$ produces an orthogonal pair  in either $c_k$ or  $d_k$. In contrast,  the states  $\KET{\phi_\pm}$ produce a ``bunched''  pair   in one of the four  detectors and are, therefore, indistinguishable. Hence, when  detectors are placed at the exit of the BS, the success probability of the BM is $50\%$~\cite{braunstein1995measurement}. 

By adding nonlinearity, one can  improve the success probability of the BM. 
To this end, we place  nonlinear routers that help distinguish between $\KET{\phi_\pm}$ at the output  of the BS [red  boxes in \figref{NL-BM}(a)].
When existing the routers, the photon pass through polarizing beam splitters (PBSs), which transmit horizontal and reflect vertical polarization, before hitting  single-photon detectors (labeled as     1-8).  When the acquired conditional phase shift in the MZIs is   $\varphi = \pi$ and when neglecting loss, all four Bell states are distinguishable by this setup. That is, each input Bell state produces a unique set of detection clicks. For example, 
only the state  $\KET{\psi_+}$  can produce clicks in~$3\,\&\,4$~or~$5\,\&\,6$ (see SM) 
\footnote{
This calculation assumes that the nonlinearity is implemented using  self-phase modulation, which takes place when two photons of the same polarization travel along the same arm of the MZI. The latter is  experimentally achievable in existing  Rydberg-EIT  platforms~\cite{firstenberg2013attractive,tiarks2019photon}.}. 

Next, we consider the effect of photon loss on our scheme, caused either by the nonlinear medium (NL loss) or by the detectors.
In the SM, we compute the success probability in the presence of loss and obtain
\begin{gather}
P_\mathrm{BM} = P_\text{DE}^2 \left[1 - \tfrac{1}{8}\left(\sqrt{1-\tau(\varphi)}\cos\varphi + 1
\right)^2 - \tfrac{\tau(\varphi)}{4}\right],
\label{eq:BM-success}
\end{gather}
where $P_\text{DE}$ denotes the single-photon detection efficiency. 
The formula is evaluated in \figref{NL-BM}(b).  
Our scheme assumes multiple detectors or  photon-number resolving detectors are used ~\cite{divochiy2008superconducting,dauler2009photon,sahin2013waveguide,Endo2021Quantum} to distinguish between instances with two photons in the same detector from instances with a single detection due to a loss event.
%~\footnote{Photon-number resolving is required in many protocols (including the linear Ewert-van Loock BM protocol discussed below) and are technologically available, e.g., by using integrable superconducting detectors that can be parallelized to make them number resolving.}.
Assuming  moderate conditional phase shift of $\varphi = \pi/3$ (achievable using available setups~\cite{liu2016large}) and $P_\text{DE} = 98\%$ improves the $P_\text{BM}$ from 0.48 to 0.67.
Evidently, in the presence of loss, the optimal operating  point is at an intermediate phase $\varphi_\text{opt}<\pi$, which  tends to $\pi$ upon increasing $\text{OD}_\text{b}$, scaling as $\pi - \varphi_\text{opt}  \propto \text{OD}_\text{b}^{-1}$ at large $\text{OD}_\text{b}$ values (Fig.~D1 in the SM). 

 \figrefbegin{NL-BM}(c)  shows our nonlinear modification of the linear Ewert-van Loock  protocol,  that  attains higher success probabilities at the cost of using two ancillary qubits~\cite{ewert20143,grice2011arbitrarily}. As shown in the SM, the success probability of the protocol is
 \begin{gather}
 P_\mathrm{EVL}  = 
 P_\text{DE}^4 \left[1 - \tfrac{1}{16}\left(\sqrt{1-\tau(\varphi)}\cos\varphi + 1
\right)^2 - \tfrac{\tau(\varphi)}{4}\right].
 \label{eq:EVL-success}
 \end{gather}
Assuming  $\pi/3$ and $P_\text{DE} = 98\%$, the success probability $P_\text{EVL}$   improves  from 0.69 to 0.77.

% ---------------------------------------------------------------------------
% FIGURE 3
% ---------------------------------------------------------------------------
\begin{figure}[t]
\centering
                 \includegraphics[width=0.5\textwidth]{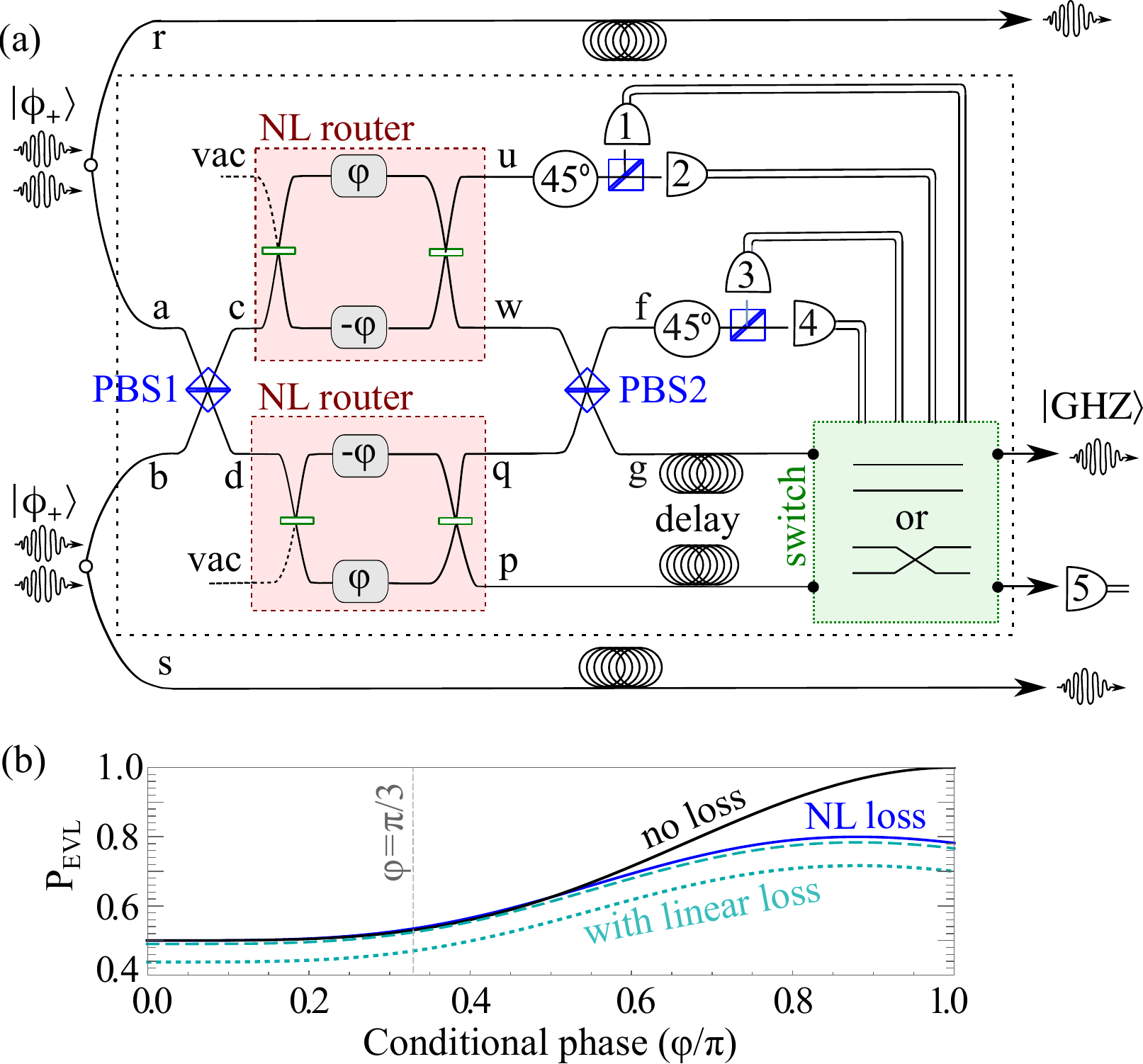}
                  \caption{\textbf{Nonlinear GHZ state generation:}  
(a) Two Bell pairs are  prepared, and   one photon  from each pair  is sent through   PBS1. Two scenarios can  produce a GHZ state:
1.~When   photons leave   PBS1 through different  ports and  exit  the routers in ports $u$ and  $p$.
2.~When    photons leave   PBS1  through the same port and also  exit the  routers  in $w$ or  $q$. 
When none of the photons are lost, zero or two clicks in detectors 1--4 or a click in 5 herald failure. 
 The switch  selects one of the output modes, $g$ or $p$, that  together with modes $r$ and $s$ contain a three-photon GHZ state when our protocol succeeds; during its operation,   photons travel in  delay lines (coils). 
(b) Success probability, $P_\text{GHZ}$, as function of $\varphi$. 
     Black solid  curve neglects loss [\eqref{GHZ-success} with $\tau=0$ and $P_\text{DE} = 1$]. Blue solid curve includes only nonlinear loss,  assuming $\text{OD}_\text{b} = 30$, while cyan dashed (dotted) curve includes also detector loss with $P_\text{DE} = 98\%$ ($90\%$). With $P_\text{DE} = 98\%$ and $\varphi = \pi/3$, $P_\text{GHZ}$ improves from $0.49$ to $0.52$.}
                  \label{fig:NL-GHZ}
  \end{figure}
% ---------------------------------------------------------------------------

% ---------------------------------------------------------------------------
\textit{Nonlinear  GHZ-state preparation:}
% ---------------------------------------------------------------------------
Our  protocol is shown  in \figref{NL-GHZ}(a). We discuss  its key features here and  provide the derivation details in the SM.  Initially, two  Bell states are prepared in $\KET{\phi_+}\!\KET{\phi_+}$ and  one  photon  from each Bell pair is sent through  a polarizing beam splitter (PBS1). Then, the photons enter nonlinear routers.  The photons may either leave  PBS1 from different ports  ($c_H^\dagger d_H^\dagger$ and $c_V^\dagger d_V^\dagger$) or through the same port ($c_H^\dagger c_V^\dagger$ and $d_H^\dagger d_V^\dagger$). 
In the former case, the photons  leave the  routers  in modes $u$ and   $p$. 
Then, the  photon in   $u$    undergoes a $45^\circ$ rotation and a subsequent measurement of the rotated photon in the diagonal basis projects the surviving photons onto a  GHZ state. This process is called ``fusion type I''~\cite{browne2005resource}. 

Our nonlinear scheme aims to ``save'' also photons that leave  PBS1 through the same port. To this end, we use nonlinear routers  that have the property that when $\varphi = \pi$,  photons from $c_H^\dagger c_V^\dagger$ and $d_H^\dagger d_V^\dagger$ are routed into modes $w_H^\dagger w_V^\dagger$ and $q_H^\dagger q_V^\dagger$.   By sending   these  photons to PBS2, rotating the photon in mode $f$, and measuring the rotated photon in the diagonal basis, a GHZ state is produced.  By adding  the probabilities for successful GHZ-state generation, either by ``fusion type I'' or by  successful nonlinear routing, we obtain (see SM)
 \begin{equation}
 P_{\mathrm{GHZ}} =P_\text{DE} \left[\frac{1}{2}+\tfrac{1}{8}\left(\sqrt{1-\tau(\varphi)}\cos\varphi - 1\right)^2\right].
 \label{eq:GHZ-success}
 \end{equation}
$ P_{\mathrm{GHZ}}$ as a function of $\varphi$  is shown in \figref{NL-GHZ}(b). 
The optimal operating point is attained at $\varphi_\text{opt}<\pi$ and scales  as  $\pi - \varphi_\text{opt} \propto \text{OD}_\text{b}^{-1/3}$ with increasing $\text{OD}_\text{b}$ (Fig.~D1). With $P_\text{DE} = 98\%$ and $\varphi = \tfrac{\pi}{3}$,  $P_\text{GHZ}$ improves from 0.49 to 0.52.

% ---------------------------------------------------------------------------
% FIGURE 4
% ---------------------------------------------------------------------------
\begin{figure}[t]
\centering
                 \includegraphics[width=0.45\textwidth]{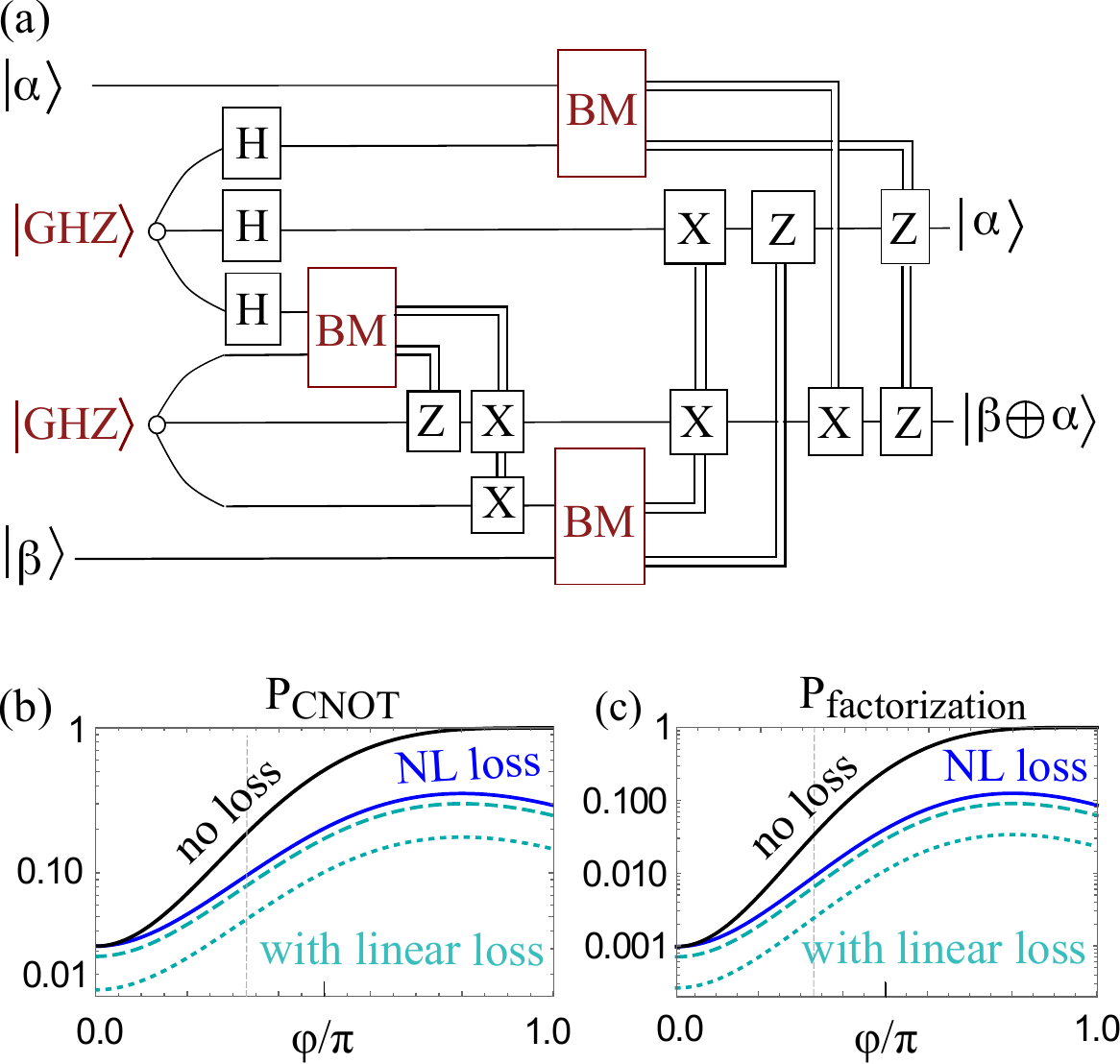}
                  \caption{\textbf{Nonlinear CNOT  and quantum factorization:}  
             (a)   The Gottesman-Chuang  linear CNOT gate~\cite{gottesman1999demonstrating}. The gate requires  three BMs, two GHZ states and   single-qubit  Pauli ($X, Z$) and Hadamard  ($H$) rotations.  (b) Success probability (log scale) of the CNOT gate, using our nonlinear protocol (with BMs and GHZs from Figs.~2-3). 
Black solid  curve neglects loss [\eqref{GHZ-success} with $\tau=0$ and $P_\text{DE} = 1$]. Blue solid curve includes only nonlinear loss,  assuming $\text{OD}_\text{b} = 30$, while cyan dashed (dotted) curve includes also detector loss with $P_\text{DE} = 98\%$ ($90\%$). 
 (c) Success probability (log scale) of quantum factorization of the number 15, using a protocol that requires   two CNOT gates~\cite{politi2009shor}, with the same building blocks and  curve labels as in  (b). For $\varphi = \pi/3$ and $P_\text{DE} = 98\%$, the success probabilities of  CNOT  and quantum factorization increase by factors of  3.32 and 11  respectively.} 
                  \label{fig:NL-GC}
  \end{figure}
% ---------------------------------------------------------------------------

Our protocol  accepts four input photons and (when successfully operated) produces three  output photons in a GHZ state. The goal of the switch is to select the modes that contain the GHZ state in order to prepare the input for a quantum computation protocol or for a subsequent stage in a cluster-state-generation protocol. 
Specifically, the switch  selects to  output mode $g$  and measure $p$ if there was a click  in detectors 3 or 4, and vice versa if  a click  occurred in detectors  1 or  2.
 During the switch operation, the   photons  travel in  delay lines.  Success of this scheme is heralded only if exactly one photon is detected - by detectors 1-4. Failure is heralded by detection of two photons in detectors 1-5 (indicating that the output is empty), or by the lack of any detection events in detectors 1-4 (whether cause by imperfect routing, detection inefficiency or loss). All these cases are accordingly disregarded.
If two photons are routed to detectors 1-5 but only one is detected due to loss (e.g. in the delay lines or the switch) or imperfect detection efficiency, this leads to a "false positive" indication of success. However, the result in this case is again the lack of photon at the output channel, and therefore is equivalent to any other loss event, and is accordingly tolerated as long as the overall loss is below the required threshold.

% ---------------------------------------------------------------------------
\textit{CNOT and Factorization:}
% ---------------------------------------------------------------------------
In  \citeasnoun{gottesman1999demonstrating}, Gottesman and Chuang (GC) present an optical-circuit implementation of  the  CNOT gate, which requires two GHZ states and three BMs [\figref{NL-GC}(a)]. Accordingly, the success probability of this protocol scales like $(P_\text{GHZ})^2 \times (P_\text{BM})^3$, being $1/2^5$ in the linear-optics case. When using our nonlinear elements, the success probability for $\varphi=\pi/3$ becomes  $0.72^3 * 0.53^2 = 0.105$, which is 3.32 larger than $1/2^5$, as shown in \figref{NL-GC}(c)~\footnote{Since finite detection efficiency limits our protocols in the same manner as the linear protocols, it does not affect the improvement factor and, hence, not included in the presented equation.}.  As CNOT is an elementary building block in most quantum protocols, this enhancement is a dramatic result. For example, the algorithm for quantum  factorization of the number 15 from~\citeasnoun{politi2009shor} requires two CNOT gates. Accordingly, our nonlinear protocols lead to an order-of-magnitude (11-fold) improvement in its success probability at $\varphi = \pi/3$ [\figref{NL-GC}(d)].

\textit{Discussion:} 
We examined photonic quantum computation protocols in the intermediate regime between linear optics and strong nonlinearity at the single photon level, and presented  efficient  protocols for key elementary operations, including BM, GHZ-state  generation, CNOT gate, and quantum factorization. 
Our results  demonstrate the potential of moderate  nonlinearity, which is achievable in a variety of platforms, using  Rydberg-EIT systems~\cite{firstenberg2013attractive,tiarks2019photon} as a relevant example. 
As photonic quantum computation, and fault-tolerant MBQC in particular, require a large number of elementary operations~\cite{li2015resource}, any modest increase in the success probability of each operation is translated to a dramatic reduction in the required resources.  For example, a conditional phase shift of $\varphi=\pi/3$, which in our scheme increases the success probability of ancilla-assisted BM with detection efficiency of 98\%  from $0.69$ to $0.77$, and of GHZ-state preparation from $0.49$ to $0.52$, can be translated into two orders of magnitude reduction in resources after  $35$ operations. 
With the recent developments in interacting atomic ensembles with integrated photonics~\cite{stern2013nanoscale,keil2016fifteen,kitching2018chip,meng2018near,corzo2019waveguide,johnson2019observation,Finkelstein2020super,pang2020hybrid}, few-photon nonlinearty on chip-scale devices is becoming feasible, making protocols that rely on moderate nonlinearities a promising new platform for photonic quantum information processing.

\textit{Acknowledgment.} The authors thank Ephraim Shahmoon,  Serge Rosenblum, and Ran Finkelstein for helpful discussions.  
AP acknowledges support of  the Koshland Foundation.  BD and OF acknowledge support from the Israeli Science Foundation.  ESK, BD,  and JPD are  also supported by  Binational Science Foundation.  ESM and JPD acknowledge the support by the U.S. Air Force Office Scientific Research, and also by the U.S. Army Research Office with the grant W911NF-
17-1-0541. BD is also supported by the Minerva Foundation, IMOD (OR RISHON),  and a research grant from Dr. Saul Unter.  OF is also supported by  the European Research Council starting investigator grant QPHOTONICS 678674.

\newpage
 \section*{Supplementary Material}
%A PARAGRAPH ON HOW TO ACHIEVE  nonlinear PHASE SHIFTS WITH ATOMS. DESCRIBE A LADDER CONFIGURATION AND EXPLAIN AC STARK SHIFT PICTURE. WITH REFERENCES TO OTHERS. 

\renewcommand{\theequation}{S\arabic{equation}}
\setcounter{equation}{0}
\numberwithin{equation}{section}
\renewcommand{\thefigure}{S\arabic{figure}}
\setcounter{figure}{0}
\numberwithin{figure}{section}

\appendix

\section*{Table of contents}
\hspace*{-0.35cm}A. Nonlinear Bell measurement\\ 
\hspace*{0.35cm}A1. Protocol without ancillas.\\
\hspace*{0.35cm}A2. Protocol with ancillas.\\
B. Nonlinear GHZ-state preparation.\\
C. Introducing single-photon phase shifts\\
\hspace*{0.35cm}C1. Nonlinear router\\
\hspace*{0.35cm}C2. Nonlinear Bell measurement.\\
\hspace*{0.35cm}C3. Nonlinear GHZ-state generation.\\
D. Selecting optimal phase shifts.

%------------------------------------------------------------------------------------------------
\section{Nonlinear Bell measurement}
\subsection{Protocol without ancillas}
%------------------------------------------------------------------------------------------------

% ---------------------------------------------------------------------------
% FIGURE A1
% ---------------------------------------------------------------------------
\begin{figure}[b]
\centering
                 \includegraphics[width=0.5\textwidth]{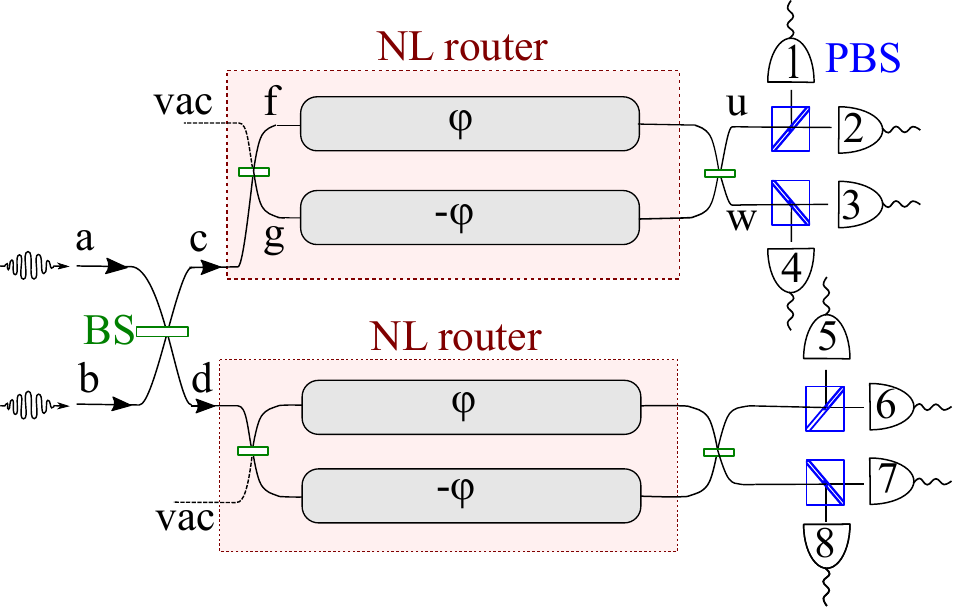}
                  \caption{For convenience, we present Fig.~2(a) from the main text, introducing labeling of the modes inside the interferometer (modes $f$, $g$, $u$ and $w$). 
}
\label{fig:BM-appendix0}
  \end{figure}
% ---------------------------------------------------------------------------

In this section, we provide   calculation details  for  the success probability of a nonlinear Bell measurement [Eq.~(8) in the main text]. For convenience, we present  Fig.~2(a) from the main text in~\figref{BM-appendix0}, and introduce additional labeling of the modes involved in the protocol.
First,  we rewrite Eq.~(7) in the diagonal basis
%\begin{subequations}
\begin{gather}
\Scale[0.9]{\KET{\psi_+} \xrightarrow[]{\text{BS}} \tfrac{i}{2\sqrt{2}}[(c_+^\dagger)^2 - (c_-^\dagger)^2  + (d_+^\dagger)^2 - (d_-^\dagger)^2  ]\KET{\text{vac}}}\nonumber\\
\Scale[0.9]{\KET{\phi_+} \xrightarrow[]{\text{BS}} \tfrac{i}{2\sqrt{2}}[(c_+^\dagger)^2 + (c_-^\dagger)^2  + (d_+^\dagger)^2 + (d_-^\dagger)^2  ]\KET{\text{vac}}}\nonumber\\
\Scale[0.9]{\KET{\phi_-} \xrightarrow[]{\text{BS}} \tfrac{i}{\sqrt{2}}[c_+^\dagger c_-^\dagger  + d_+^\dagger d_-^\dagger  ]\KET{\text{vac}}}.
\label{eq:Bell-states-diagonal}
\end{gather}
%\end{subequations}
To trace the evolution upon entering and leaving the MZI, we invoke the beam-splitter (BS) transformation
\begin{gather}
c^\dagger\rightarrow\tfrac{1}{\sqrt{2}}(f^\dagger+ig^\dagger)\nonumber\\
f^\dagger\rightarrow\tfrac{1}{\sqrt{2}}(w^\dagger+iu^\dagger)\nonumber\\
g^\dagger\rightarrow\tfrac{1}{\sqrt{2}}(u^\dagger+iw^\dagger).
\end{gather}
Due to self-phase modulation in the atomic medium,  an  identical photon pair that enters modes $f$ or $g$ undergoes the  transformation 
\begin{gather}
(f_\pm^\dagger)^2 \rightarrow \sqrt{1 - \tau}\,e^{i\varphi}(f_\pm^\dagger)^2   +
\sqrt{\tau} f_\pm^\dagger \ell_\pm^\dagger
\\
(g_\pm^\dagger)^2 \rightarrow \sqrt{1 - \tau}\,e^{i\varphi}(g_\pm^\dagger)^2   +
\sqrt{\tau} g_\pm^\dagger k_\pm^\dagger,
\label{eq:self-phase-rules}
\end{gather}
where the first term on the right hand side accounts for phase acquisition and the second for loss. 
By using \eqssref{Bell-states-diagonal}{self-phase-rules}, we find 
\begin{gather}
\Scale[0.9]{(c_\pm^\dagger)^2   \rightarrow
\tfrac{\sqrt{1-\tau}\cos\varphi-1}{2}
(w_\pm^\dagger)^2  -
\tfrac{\sqrt{1-\tau}\cos\varphi+1}{2}
(u_\pm^\dagger)^2   -}
\nonumber\\
\Scale[0.9]{
 \sqrt{1-\tau} \sin\varphi \,w_\pm^\dagger u_\pm^\dagger +
\tfrac{\sqrt{\tau}}{2\sqrt{2}}
  \left[(w_\pm^\dagger+iu_\pm^\dagger)l_\pm^\dagger - (u_\pm^\dagger+iw_\pm^\dagger)k_\pm^\dagger\right],
  }
  \label{eq:ref-in-main-text}
\end{gather}
generalizing Eq.~(6) in the main text. A similar transformation rule applies for the $d_+d_-$ component of the wavefunction. 

Both the states $\KET{\phi_+}$ and $\KET{\psi_+}$ are affected by the nonlinearity, since both contain identical photon pairs [see~\eqref{Bell-states-diagonal}]. 
However, whenever two photons are detected, $\KET{\phi_+}$ is distinguishable from the remaining states because it is the only state that produces a pair of orthogonal photons either in the upper detectors (1--4) or in the lower detectors  (5--8). Therefore, when both photons reach the detectors, only the cases where both photons from $\KET{\psi_+}$ are routed into $u_+^\dagger u_-^\dagger$ lead to failure. 
Therefore, the failure probability is found by collecting the terms proportional to $u_+^\dagger u_-^\dagger$. 
Photon loss also leads to failure of the BM. 
Both $\KET{\phi_+}$ and $\KET{\psi_+}$ can experience loss, and the final states in that case are indistinguishable.  Given that the initial state is $\KET{\psi_+}$ or $\KET{\phi_+}$, the probability for loss is $\tau/2$. Therefore, the overall success probability becomes
\begin{gather}
P_\mathrm{BM} = P_\text{DE}^2
\left[1 - \tfrac{1}{8}\left(\sqrt{1-\tau(\varphi)}\cos\varphi + 1\right)^2 - \tfrac{\tau(\varphi)}{4}\right],
\end{gather}
where the last term corresponds to nonlinear-loss events and $P_\text{DE}$ accounts for finite detection efficiency.

% ---------------------------------------------------------------------------
\subsection{Protocols with ancillas  }
% ---------------------------------------------------------------------------

\begin{figure}[t]
\centering
                 \includegraphics[width=0.5\textwidth]{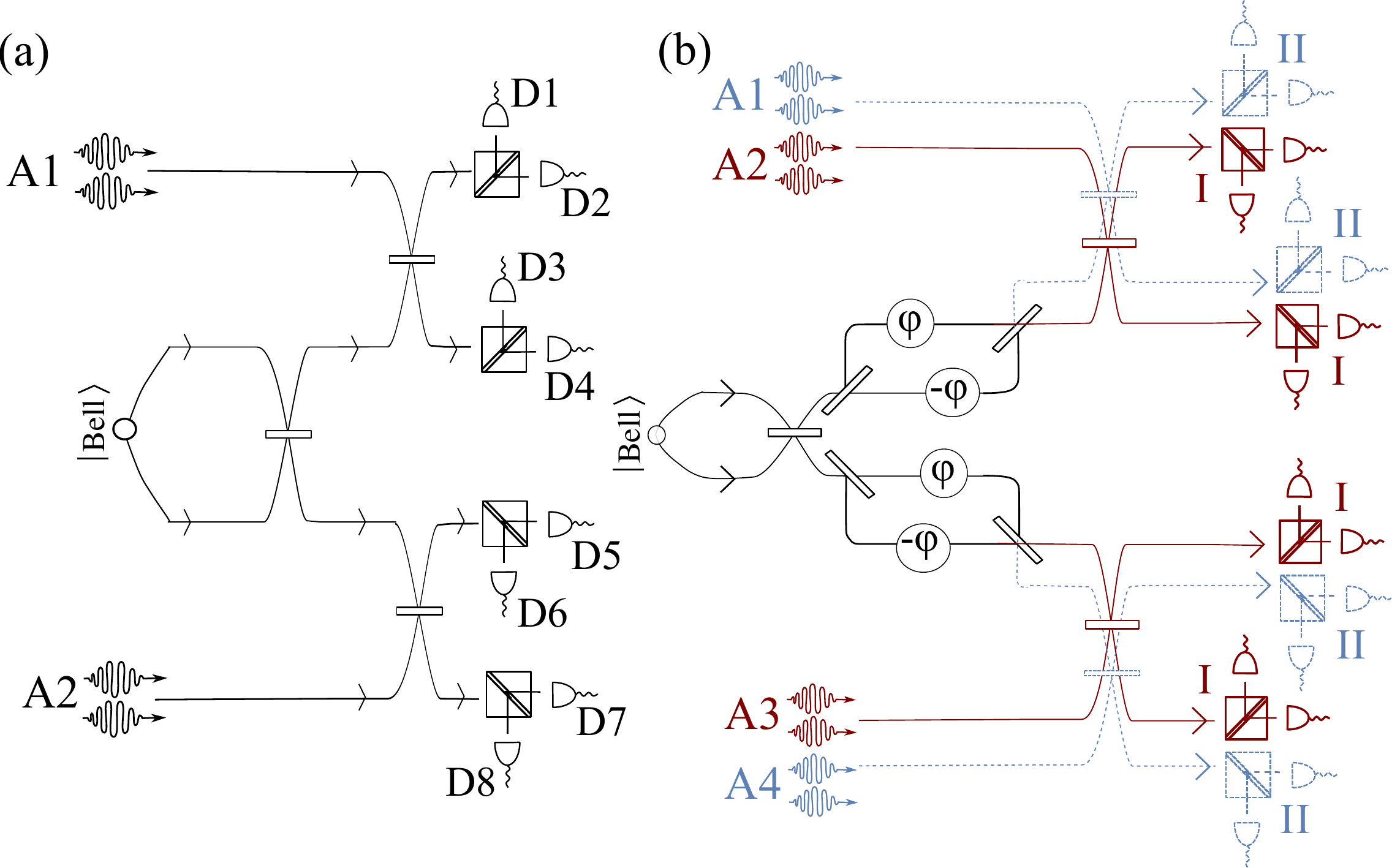}
                  \caption{(a) Bell-state measurement using four ancillary qubits 
                  prepared in $ \tfrac{1}{2}[{(a^\dagger_{H})^2 +  (a^\dagger_{V})^2}]\KET{\text{vac}}$. 
                 A Bell state is sent through the device, and the final measurement can determine the input state with success probability of $75\%$. (b) By introducing nonlinear routers (and additional ancillas and detectors), the success probability increases, and reaches $100\%$ for $\pi$ conditional  phase shifts, as shown by the green line in Fig. 2(b) in the main text.}
                  \label{fig:ancilla-based-BM}
  \end{figure}

In the main text, we present a Bell measurement (BM) that uses conditional phase shifts to  improve a linear method whose success probability is $50\%$.  However,  linear protocols that  use ancillary photons may  have higher success rates, with probability approaching $100\%$; the failure rate drops exponentially with the  number of ancillary photons~\cite{grice2011arbitrarily}. Fig.~2(c) depicts the success rate of a refined nonlinear protocol, which is based on a linear method  with  two ancillary qubits and   success probability of $75\%$~\cite{ewert20143}. The linear protocol and its nonlinear improvement are  shown in \figref{ancilla-based-BM}(a) and (b) respectively.

% ---------------------------------------------------------------------------
% FIGURE A2
% ---------------------------------------------------------------------------
% ---------------------------------------------------------------------------

The linear method, proposed independently by Grice and Ewert and van-Loock~\cite{grice2011arbitrarily,ewert20143}, utilizes  additional photons prepared in $\KET{A_{i}} = \tfrac{1}{2}[{(a^\dagger_{i,H})^2 +  (a^\dagger_{i,V})^2}]\KET{\text{vac}}, $ where the index $i = 1,2$ enumerates the ancillas and $\KET{\text{vac}}$ is the vacuum of the acillary modes.  The protocol is shown in\figref{ancilla-based-BM}(a). we briefly revise the linear method before introducing our nonlinear modification. By using 8 detectors, the Bell state can be determined with success probability of $75\%$. The states   $\psi_{\pm}$ and $\phi_{\pm}$ differ in the parity of $H$ and $V$ polarized photons, since the parity is unaltered by the device. The states $\psi_{\pm}$ can be distinguished since they differ in the number of photons that reach the upper detectors ($D1,\hdots,D4$). Finally, the states $\phi_{\pm}$  can only be distinguished with probability $50\%$ (instances where all detected photons have the same polarization can occur for both $\phi_{+}$ and $\phi_{-}$ and lead to failure). 

By introducing two nonlinear routers, one in each output of the first beam splitter, we increase the probability to distinguish between the $\phi_{\pm}$ states. The setup is shown in \figref{ancilla-based-BM}(b). The success probability of our nonlinear protocol (when neglecting loss) is
 \begin{gather}
 P_\mathrm{EVL}  = 
 P_\text{DE}^4 \left[1 - \tfrac{1}{16}\left(\sqrt{1-\tau(\varphi)}\cos\varphi + 1
\right)^2 - \tfrac{\tau(\varphi)}{4}\right].
 \end{gather}
With a conditional phase shift of $\pi$, the BM becomes deterministic. Note, however, that  the nonlinear setup requires four ancillary qubits and eight detectors (while the linear counterpart required only two ancilla qubits). 

% ---------------------------------------------------------------------------
\section{Nonlinear GHZ-state preparation }
% ---------------------------------------------------------------------------

% ---------------------------------------------------------------------------
% FIGURE B1
% ---------------------------------------------------------------------------
\begin{figure}[b]
\centering
                 \includegraphics[width=0.5\textwidth]{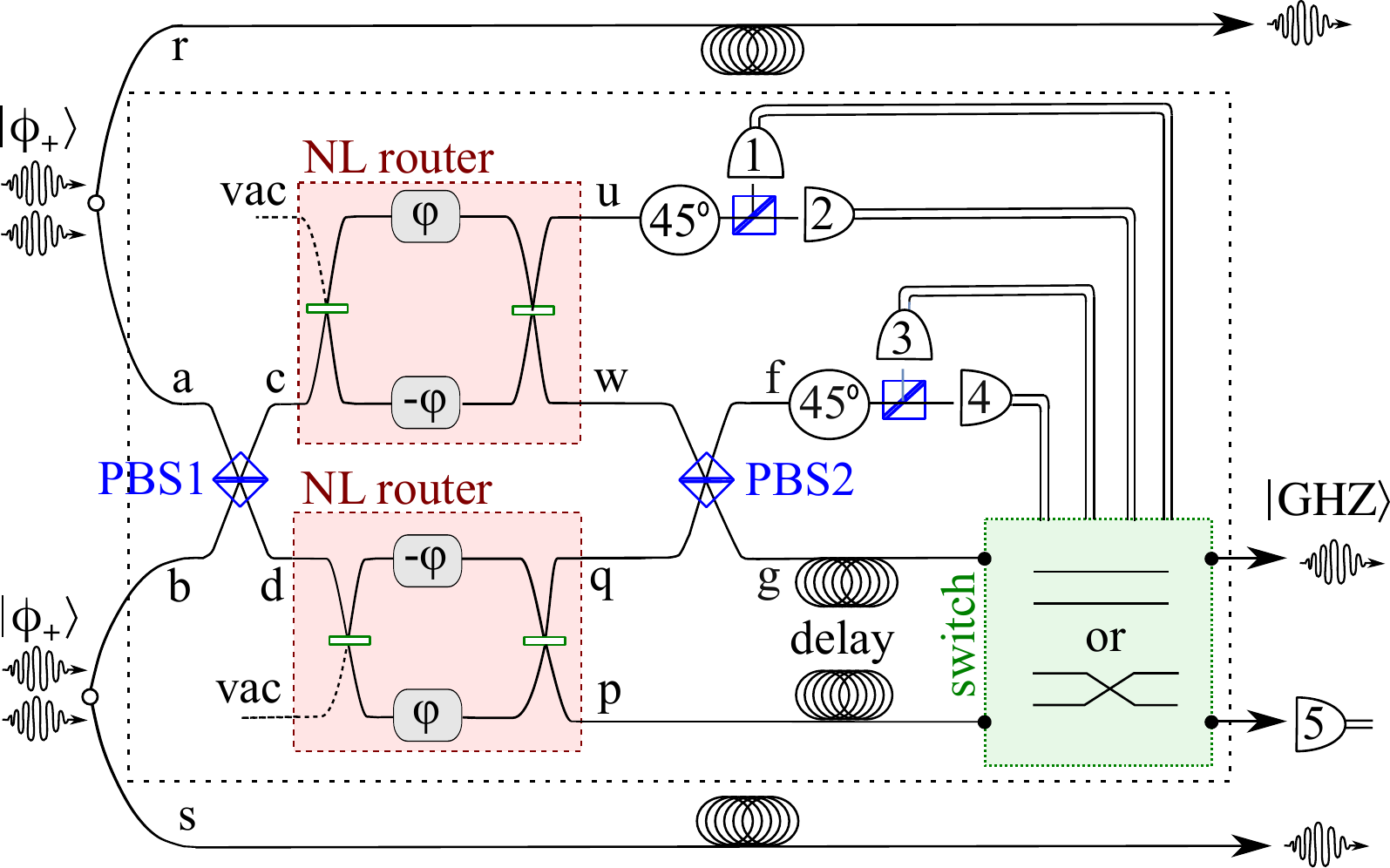}
                  \caption{To fascillitate the reading of  the derivation, we present Fig.~3(a) from the main text also here. 
}
\label{fig:GHZ-appendix}
  \end{figure}
% ---------------------------------------------------------------------------

In the main text, we describe a nonlinear protocol for GHZ-state generation.  In this section, we provide the details of the derivation. 
For the the notation we refer to~\figref{GHZ-appendix}.
The protocol  uses a resource of two  Bell pairs, 
 \begin{gather}
 \KET{\psi} = \tfrac{1}{2}(r_H^\dagger a_H^\dagger + r_V^\dagger a_V^\dagger)(s_H^\dagger   b_H^\dagger + s_V^\dagger  b_V^\dagger)\KET{\text{vac}}.
 \end{gather}
 Then, one photon from each pair is sent through  a polarizing beam splitter (PBS) (which transmits $H$-polarized photons and reflects $V$ photons) and the state becomes  
 \begin{gather}
 \KET{\psi} \xrightarrow[]{\text{PBS1}} \tfrac{1}{2}(r_H^\dagger d_H^\dagger +i r_V^\dagger c_V^\dagger)(s_H^\dagger   c_H^\dagger + i s_V^\dagger  d_V^\dagger)\KET{\text{vac}}.
 \end{gather}
This wavefunction contains two types of components: 
ones   with single photons in each output  arm of the PBS   ($c_H^\dagger d_H^\dagger$ and $c_V^\dagger d_V^\dagger$) and others  with both photons in the same arm ($c_H^\dagger c_V^\dagger$ and $d_H^\dagger d_V^\dagger$).
The former type of terms lead to successful GHZ-state generation. For these states, our protocol reproduces the linear protocol ``fission I.''
For these terms, the photons  leave the routers  through  ports  $u$ or  $p$.  
Then, the photon in   $u$    undergoes a $45^\circ$ rotation, resulting in 
  \begin{gather}
  \Scale[0.9]{
 \tfrac{1}{2}(r_H^\dagger s_H^\dagger  c_H^\dagger d_H^\dagger  - r_V^\dagger s_V^\dagger   c_V^\dagger d_V^\dagger)\xrightarrow[]{\text{MZI}}
 \tfrac{1}{2}(r_H^\dagger s_H^\dagger  p_H^\dagger u_H^\dagger  - r_V^\dagger s_V^\dagger   p_V^\dagger u_V^\dagger)
 }\nonumber\\
 \Scale[0.9]{\xrightarrow[]{45^\circ}
 \tfrac{1}{2\sqrt{2}}[ u_+^\dagger( r_H^\dagger s_H^\dagger  p_H^\dagger  - r_V^\dagger s_V^\dagger  p_V^\dagger )  + u_-^\dagger
 ( r_H^\dagger s_H^\dagger  p_H^\dagger  + r_V^\dagger s_V^\dagger  p_V^\dagger )].}
 \end{gather}
Therefore, a click in either $u_+$ or $u_-$ projects the surviving photons onto a GHZ state.

On the other hand, our protocol can save also the latter type of terms. 
The wavefunction components with  both photons  in the same arm (containing $c_H^\dagger c_V^\dagger$ and $d_H^\dagger d_V^\dagger$)  are routed into modes $u, p$ or $w, q$ with probabilities shown in~Fig.~1(c). For these terms, the cases where both photons are also routed into $w$ or $q$ lead to successful GHZ generation. 
This is because the latter transform as
  \begin{gather}
\Scale[0.9]
{ \tfrac{1}{2}(r_V^\dagger s_H^\dagger  w_H^\dagger w_V^\dagger  + r_H^\dagger s_V^\dagger   q_H^\dagger q_V^\dagger)
\xrightarrow[]{\text{PBS2}}
\tfrac{i}{2}(r_V^\dagger s_H^\dagger  g_H^\dagger f_V^\dagger  + r_H^\dagger s_V^\dagger   f_H^\dagger g_V^\dagger)}
\nonumber\\
\Scale[0.9]{
\xrightarrow[]{45^\circ} \tfrac{i}{2\sqrt{2}}[f_+^\dagger(r_V^\dagger s_H^\dagger  g_H^\dagger  + r_H^\dagger s_V^\dagger    g_V^\dagger)+
 f_-^\dagger(r_V^\dagger s_H^\dagger  g_H^\dagger  - r_H^\dagger s_V^\dagger    g_V^\dagger)]}.
 \label{eq:last-equation}
 \end{gather}
 By scrutinizing the terms in the second line of~\eqref{last-equation}, one can see that a measurement of the photon in mode $f$ in the diagonal basis projects the surviving photons onto a GHZ state. 

%---------------------------------------------------------------------------
\section{ Introducing single-photon phase shifts}
%---------------------------------------------------------------------------

% ---------------------------------------------------------------------------
% FIGURE 2 Appendix
% ---------------------------------------------------------------------------
\begin{figure*}[t]
\centering
                 \includegraphics[width=1\textwidth]{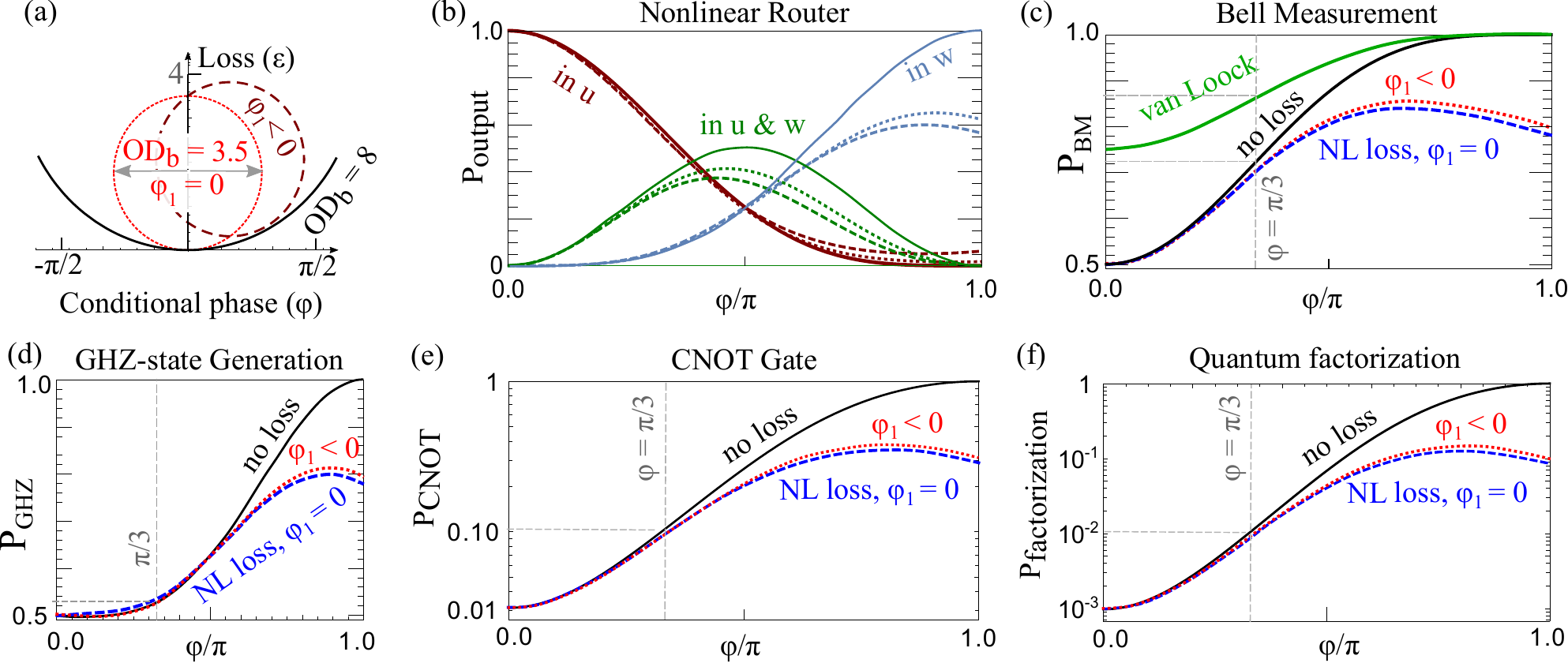}
                  \caption{\textbf{Adding a single-photon phase shift, $\mathbf{\varphi_1}$, to reduce the nonlinear loss:} 
(a) Conditional phase-loss relation in Rydberg EIT systems, using \eqref{phase-loss-phi1}. Same as Fig.~1(c), with the addition of the case $\varphi_1 = -0.5, \mathrm{OD}_\mathrm{b} = 3.5$ (brown-dashed curve). 
(b) Nonlinear router: Probabilities for detection outcomes, $P_\text{output}$, for an incoming photon pair. Same as in Fig.~1(b),  with the addition of the case $\varphi_1 = - \tfrac{\varphi}{11}$, evaluated using \eqref{MZI-with-loss-append} (dotted curves here and in all of the following subplots in this figure). This value of $\varphi_1$ is  chosen to optimize the success probability and is used in all remaining subplots. 
(c--f) Success probabilities of a nonlinear BM ($P_\text{BM}$), GHZ-states generation ($P_\text{GHZ}$), CNOT gate ($P_\text{CNOT}$), and quantum factorization ($P_\text{factorization}$) as a function of $\varphi$. Same as Figs.~2(c), 3(b), 4(b,c), with the addition of the case $\varphi_1<0$.}
                  \label{fig:phi1-append}
  \end{figure*}
% ---------------------------------------------------------------------------

%---------------------------------------------------------------------------
\subsection{Nonlinear router }
%---------------------------------------------------------------------------
In this appendix, we introduce single-photon phase shifts in order to achieve conditional (two-photon) phase shifts with reduced nonlinear loss. 
In the applications that we study in this paper, we find that this trade-off between single-photon phase shifts and loss leads to a  minor improvement in the performance.  The results of our analysis are shown in \figref{phi1-append}.

In order to reduce the nonlinear loss, we choose photon frequencies that are detuned from the EIT resonance, such that a photon traveling through the medium acquires a  phase shift $\varphi_1$ and experiences loss, described by the absorption coefficient  $\varepsilon_1$.
Once a photon generates  a Rydberg excitation, a subsequent photon inside the blockade radius  effectively experiences an ensemble of two-level atoms and acquires a phase shift   $\varphi_2$ and loss with absorption coefficient $\varepsilon_2$. When neglecting loss, this model produces the following transformation rules for the creation operators in the MZI [Fig.~1(a) in the main text]:
\begin{gather}
f^\dagger\rightarrow e^{i\varphi_1}f^\dagger \quad,\quad  {f^\dagger}^2\rightarrow e^{i(\varphi_1+\varphi_2)}{f^\dagger}^2 \\
 g^\dagger\rightarrow e^{-i\varphi_1}g^\dagger
\quad,\quad {g^\dagger}^2\rightarrow e^{-i(\varphi_1+\varphi_2)}{g^\dagger}^2.
\label{eq:transformation-nonzero-phi1}
\end{gather}
Our motivation is that by introducing small negative $\varphi_1$, one needs a 
smaller $\varphi_2$ to achieve the same conditional phase shift, 
\begin{gather}
\varphi \equiv \varphi_2 - \varphi_1 .
\label{eq:nonlinear-phase-def}
\end{gather}
Since the phase shifts and loss coefficients satisfy the circle relations [\eqref{circle-append} below], we expect to reduce the total  loss
\begin{gather}
\varepsilon\equiv  \varepsilon_1 + \varepsilon_2. 
\end{gather}

Once  introducing  $\varphi_1\neq0$, one must balance the MZI (by adding a linear phase shift to one of its arm) in order for single photons to exit deterministically through one port. Revisiting Eq.~(1) from the main text and introducing single-photon phase shifts, $e^{\pm\varphi_1}$, and an additional phase shift of $e^{i\Delta}$ for mode $f$, we obtain
 \begin{gather}
a^\dagger \xrightarrow[]{\text{BS1}} \tfrac{1}{\sqrt{2}}(e^{i(\varphi_1+\Delta)}f^\dagger + ie^{-i\varphi_1}g^\dagger)\xrightarrow[]{\Delta = -2\varphi_1}\nonumber\\
\tfrac{ e^{-i\varphi_1}}{\sqrt{2}}
(f^\dagger + i g^\dagger) \xrightarrow[]{\text{BS2}} e^{-i\varphi_1} u^\dagger.
 \end{gather}
With this choice of $\Delta$, the MZI is balanced.  
Next, let us trace the propagation of a photon pair in the MZI.  Revisiting Eq.~(3), we find
\begin{gather}
\Scale[0.9]{
\left(a^\dagger\right)^2 \!\! \xrightarrow[]{\text{BS1}}\! 
\tfrac{1}{2}\left(e^{i\tfrac{\varphi_1 + \varphi_2+\Delta}{2}}f^\dagger + ie^{-i\tfrac{\varphi_1+\varphi_2}{2}}g^\dagger\right)^2
\xrightarrow[]{\Delta = -2\varphi_1}}\nonumber\\\Scale[0.9]{ 
\tfrac{e^{-2i\varphi_1}}{2}\left(e^{i\tfrac{ \varphi}{2}}f^\dagger + ie^{-i\tfrac{\varphi}{2}}g^\dagger\right)^2
\xrightarrow[]{\text{BS2}} 
e^{-2i\varphi_1} \left(w^\dagger \sin\tfrac{\varphi}{2} + u^\dagger 
  \cos\tfrac{\varphi}{2}\right)^2}.
\end{gather}
In the second line, we used the definition of $\varphi$ [\eqref{nonlinear-phase-def}].

Next, we account for  the effect of loss on the probability amplitude of the surviving terms. For brevity, we do not keep track of the loss channels and denote them symbolically by the word ``loss'' in Eqs.~(C7-11) below. The transformation rule [\eqref{transformation-nonzero-phi1}] is modified:
\begin{gather}
f^\dagger\rightarrow \sqrt{1-\tau_1}e^{i\varphi_1}f^\dagger +\text{loss},\\
g^\dagger\rightarrow \sqrt{1-\tau_1}e^{-i\varphi_1}g^\dagger + \text{loss},\\
{f^\dagger}^2\rightarrow \sqrt{(1-\tau_1)(1-\tau_2)}e^{i(\varphi_1+\varphi_2)}{f^\dagger}^2 + \text{loss},
\\
{g^\dagger}^2\rightarrow \sqrt{(1-\tau_1)(1-\tau_2)}e^{-i(\varphi_1+\varphi_2)}{g^\dagger}^2 + \text{loss}.
\end{gather}
Tracing the propagation of photon pairs through the MZI, we find [generalizing  Eq.~(5) in the main text]
\begin{gather}
\Scale[0.9]{{a^\dagger}^2   \rightarrow
\tfrac{\sqrt{(1-\tau_1)(1-\tau_2)}}{2}e^{-2i\varphi_1}\left[\cos\varphi
\left({w^\dagger}^2 \! -{u^\dagger}^2  \right) - 2\sin\varphi (w^\dagger u^\dagger) \right]}
\nonumber\\
\Scale[0.9]{
\!-\tfrac{(1-\tau_1)}{2}e^{-2i\varphi_1}
\left({w^\dagger}^2  +{u^\dagger}^2\right)+\text{loss}}.
  \label{eq:MZI-with-loss-append}
\end{gather}
The acquired phase shifts and loss coefficients satisfy the circle relation:
\begin{gather}
\varphi_i^2 + \left(  \tfrac{\text{OD}}{4} - \tfrac{\varepsilon_i}{2}\right)^2 = 
\left(  \tfrac{\text{OD}}{4}\right)^2,
\label{eq:circle-append}
\end{gather}
for $i = 1,2$, where $e^{-\varepsilon_i} = 1-\tau_i$.

\eqrefbegin{MZI-with-loss-append} is used in Fig.~1(a) in the main text to compute the probability for a photon pair to exit in modes $w, u$ or both when $\varphi_1\neq0$ (dotted curves). The probabilities can be expressed in terms of $\varphi_1, \varphi, \varepsilon_1,$ and $\varepsilon$. We choose $\varphi_1$ and $\varphi$ and express $\varepsilon_1(\varphi_1)$  using \eqref{circle-append} and the overall loss,
using 
\begin{gather}
\varepsilon(\varphi,\varphi_1) = \varepsilon_1(\varphi_1) + \tfrac{\text{OD}}{2} \pm \sqrt{\left(  \tfrac{\text{OD}}{2}\right)^2 - 4(\varphi + \varphi_1)^2}.
\label{eq:phase-loss-phi1}
\end{gather}

%------------------------------------------------------------------------
\subsection{Nonlinear Bell measurement}
%------------------------------------------------------------------------

Introducing conditional phase shifts  for identical photon pairs, the transformation rule for creation operators becomes
\begin{gather}
{f_\pm^\dagger}^2   \rightarrow \sqrt{(1-\tau_1)(1-\tau_2)}e^{i(\varphi_1+\varphi_2)}{f_\pm^\dagger}^2    +\text{loss},\\
f_+^\dagger f_-^\dagger
 \rightarrow (1-\tau_1) e^{2i\varphi_1}f_+^\dagger f_-^\dagger  +\text{loss}.
\end{gather}
Recall that the states $\KET{\psi_-}$ and $\KET{\phi_-}$ do not contain identical photon pairs in the diagonal basis after the first BS   [see \eqref{Bell-states-diagonal}]. Therefore, these states can only encounter single-photon loss and  the survival probability for each of these states is $(1-\tau_1)^2$. Conversely, the states $\KET{\psi_+}$ and $\KET{\phi_+}$ contain identical photon pairs. By summing the probability of the no-loss terms in \eqref{MZI-with-loss-append}, one finds that the survival probability for each of these states is $\frac{1}{2}[(1-\tau_1)^2  + (1-\tau_1)(1-\tau_2)]$. The interpretation of the last result is that either the photons survive two single-photon loss events (in an EIT medium) or they survive a single photon and a second-photon loss (in an effective two-level medium) event. 
The overall survival probability is
\begin{gather}
P_\mathrm{survive} = \tfrac{(1-\tau_1)^2}{2}+\tfrac{(1-\tau_1)(2-\tau_1-\tau_2)}{4}.
\end{gather}
Therefore, the success probability of a BM  is 
\begin{gather}
\Scale[0.9]{
P_\mathrm{BM} =P_\mathrm{survive} - \tfrac{1}{8}\left(
\sqrt{(1-\tau_1)(1-\tau_2)}\cos\varphi + (1-\tau_1)
\right)^2}.
\label{eq:BM-self-with-1phot-loss}
\end{gather}
To account for finite detection efficiency, one needs to multiply this result by $P_\text{DE}^2$.

%------------------------------------------------------------------------
\subsection{Nonlinear GHZ-state generation}
%------------------------------------------------------------------------
To compute the success probability of our protocol in the presence of $\varphi_1\neq0$, recall that there are two scenarios that produce a GHZ state: 
(i) When the photon pair leaves PBS1 through different ports and both photons reach the final detectors (avoiding absorption due to linear loss in the MZI), and  
(ii) When the photons leave PBS1 through the same port and, later, also leave the MZI in modes $w$ or $q$.
Adding  the probabilities for these scenarios, we obtain
\begin{gather}
\Scale[0.9]{
P_\mathrm{GHZ} = \tfrac{1}{2}(1-\tau_1)^2
- \tfrac{1}{8}\left(\sqrt{(1-\tau_1)(1-\tau_2)}\cos\varphi + (1-\tau_1)\right)^2}.
\label{eq:GHZ-with-1phot-loss}
\end{gather}

%------------------------------------------------------------------------
\section{Selecting optimal phase shifts}
%------------------------------------------------------------------------

% ---------------------------------------------------------------------------
% FIGURE 1 Appendix
% ---------------------------------------------------------------------------
\begin{figure}[t]
\centering
                 \includegraphics[width=0.5\textwidth]{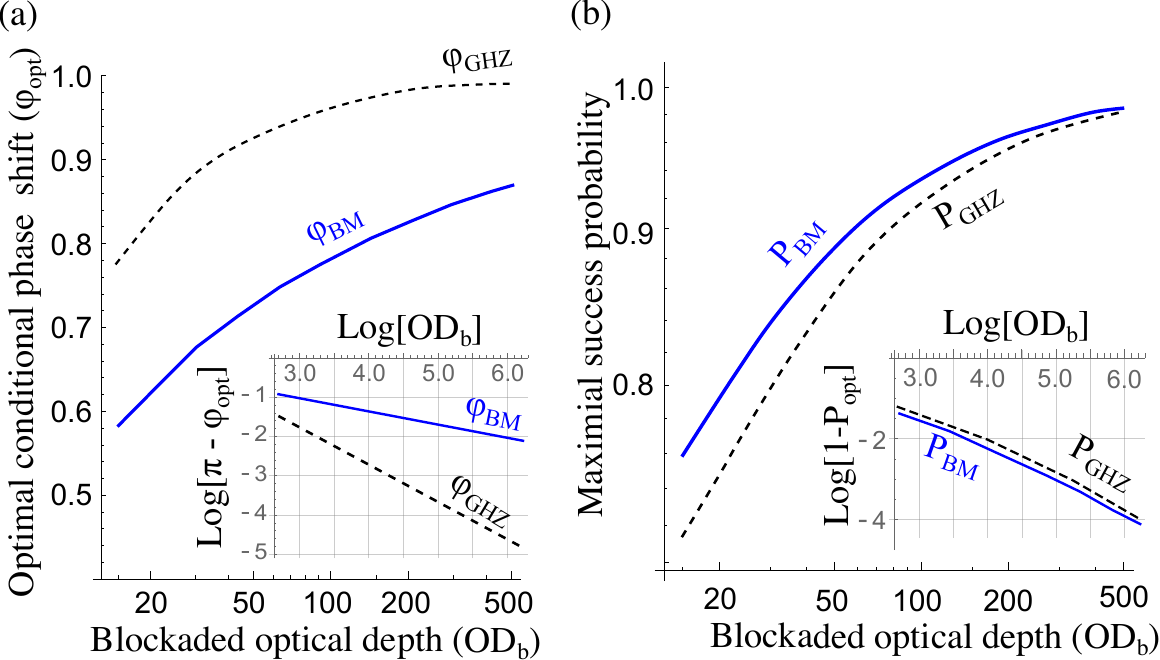}
                  \caption{(a) Optimal phase shift  as a function of the blockaded optical depth  for BMs (blue solid) and for GHZ-state generation (black dashed). (b) The success probabilities at the optimal phases shown in (a). The insets show the same data on a logarithmic scale.}
                  \label{fig:optimal-phase-shifts}
  \end{figure}
% ---------------------------------------------------------------------------
When examining  the plots in the main text that show of the success probability versus $\varphi$,   a surprising feature emerges: While one would naively expect that $\pi$ phase shifts yield optimal results, it turns out that in all studied applications, the optimal phase is smaller than $\pi$. In~\figref{optimal-phase-shifts}(a), we compute the optimal phase shift, $\varphi_\text{opt}$, as a function of the blockaded optical  depth, OD$_b$, for or  BM  and  GHZ-state generation protocols. \figrefbegin{optimal-phase-shifts}(b) shows the corresponding success probabilities at the optimal phase shifts from (a). The insets show the same data as the main plots on a logarithmic scale. One can see that at large optical depths ($\text{OD}_\text{b}>50$), all curves in the insets are linear. We find that the optimal phases scale as $\varphi_\text{opt} - \pi \propto \mathrm{OD}^{\alpha}$, where $\alpha = -1$ for the GHZ-state preparation protocol and $\alpha = -0.3$ for BMs. The infidelity of our protocols scales as $1 - P_\text{opt}  \propto \mathrm{OD}^{\alpha}$, where  $\alpha \approx  -0.9$ for both protocols.

\bibliography{WEAKbibli}

%\section{COMMENTS}
%Estimate experimentally achievable nonlinear phases.  Show how this translates to by how much the overhead in single photons is reduced. 

\end{document}